**3D confinement of water: $H_2O$ exhibits long-range (> 50 nm) structure while $D_2O$ does not**

N. Dupertuis[1], O. B. Tarun[1], C. Lütgebaucks[1], and S. Roke[1]*

[1]Laboratory for Fundamental BioPhotonics (LBP), Institute of Bioengineering (IBI), and Institute of Materials Science (IMX), School of Engineering (STI), and Lausanne Centre for Ultrafast Science (LACUS), École Polytechnique Fédérale de Lausanne (EPFL), CH-1015 Lausanne, Switzerland,
e-mail*: sylvie.roke@epfl.ch

**Abstract**

Water is the liquid of life, thanks to its three-dimensional adaptive hydrogen (H)-bond network. Confinement of this network may lead to dramatic structural changes that influence chemical and physical transformations. Although confinement effects occur on a < 1 nm length scale, the upper length scale limit is not known. Here, we investigate 3D confinement over lengths scales ranging from 58 -140 nm. By confining water in zwitterionic liposomes of different sizes and measuring the change in H-bond network conformation using second harmonic scattering (SHS) we determined long range confinement effects in light and heavy water. $D_2O$ displays no detectable 3D-confinement effects < 58 nm (< $3·10^6$ $D_2O$ molecules). $H_2O$ is distinctly different: The vesicle enclosed inner H-bond network has a different conformation compared to the outside network and the SHS response scales with the volume of the confining space. $H_2O$ displays confinement effects over distances >140 nm (>$4·10^7$ $H_2O$ molecules).

## Introduction

As the primary solvent of life, water is a fascinating liquid. Liquid water is structured in a three-dimensional dynamic network that is highly adaptive to its surroundings[1]. Interfaces[2] are able to impose a structure on water, electric fields align water molecules[3,4], and macromolecules[5,6] such as DNA[7,8] and aquaporin[9] can impose chirality on the water molecules adjacent to it. Furthermore, nanopores may induce freezing[10,11], and water in between graphene sheets displays non-bulk like behavior that is presently not understood[12,13]. Thus, confinement – the restriction of available space – influences water structure. But the way in which it does is not well understood: Are confinement effects, purely stemming from local interactions, such as ion-dipole, van der Waals and direct hydrogen (H)-bonding interactions, or are collective, entropic effects involving many water molecules also important? Likewise, what is the length scale over which confinement plays a role? If a collective / long range effects play a role, we may expect the length scale for confinement effects to greatly exceed 1 nm.

Currently available experimental and theoretical studies agree that the length scale over which confinement effects occur is up to a few nanometers: Vibrational dynamics studies of water in reverse micelles that are a few nanometers in size show a non-bulk like vibrational relaxation[14–18]. Carbon nanotubes modify water dynamics and thermodynamics compared to bulk water[10,19–21]. Freezing transitions are observed in pores with diameters < 2 nm[22–24]. Besides such small systems, structural changes have been probed on macroscopic extended planar interfaces, using atomic and surface force microscopies, whereby the water is confined between two macroscopic surfaces, or between an AFM tip and an extended planar interface. These studies show structural changes at separations of only a few water layers[25–28], and these modifications are explained in terms of water-interface[29] interactions.

However, collective entropic interactions are potentially very important: As one confines water the number of H-bond network configuration changes[18], which is for example very clearly seen in the infrared spectra of the gas to liquid phase transition of small water clusters[30,31]. That these have not been reported yet, is caused by the fact that probing interactions that exceed several nanometers under 3-dimensional (3D) confinement is a great challenge. Most experimental techniques that probe the structure of water, such as dielectric and THz spectroscopy[32–34], X-ray and neutron scattering[5,35,36], vibrational spectroscopies[37,38], vibrational dynamics studies and NMR spectroscopy[17,19,36,39,40] are sensitive to local, ≤ 1 nm perturbations to the H-bond network of water. Likewise, computational methods on the quantum chemistry level have been performed over length scales smaller than a few nanometers[24,41–44]. Additionally, most confinement experiments that would allow for longer length scale effects, have been done using partial 1D and 2D confinement[39] in combination with time scales that are much larger than the reorganization time of water. Therefore, it is possible that equilibration of the aqueous system obscures longer-range, collective confinement effects.

Here, we induce 3D confinement by encapsulation light and heavy water using zwitterionic overall charge-neutral liposomes, with diameters in the range from 58 to 140 nm. We compare the resulting changes in water structure to the water structure adjacent to a planar extended 2D lipid bilayer made of the same lipids. To determine structural changes to the H-bond network of water, we quantify the structural difference between the inside and outside water using femtosecond (fs) second harmonic (SH) imaging and angle resolved[45] (AR) second harmonic scattering (SHS). Non-confined water adjacent to a planar bilayer produces a vanishing SH signal, since there is no difference between the water adjacent to the top and bottom leaflets. In contrast, all measured liposomes display differences in the water structure and the size and number density normalized response increases with decreasing size. Interestingly, the SHS water response of $D_2O$ is significantly (~8 x) smaller than that of $H_2O$ and the size dependence is different. Whereas the structural changes in $D_2O$ are explained by an obvious increase in the relative amount of interfacial water per liposome, the structural changes in $H_2O$ depend on the available confined volume and are therefore due to 3D confinement of the hydrogen bond network. Therefore, light water exhibits long range structural changes over distances up to 140 nm, while heavy water does not up to 56 nm, which is the size limit of our experiments.

**Results and Discussion**

**Quantifying confinement effects in water.** To determine if structural changes occur under 3D confinement, we compare the structure of water in contact with a free-standing lipid bilayer with water in contact with different sized liposomes composed of the same bilayers. We quantify the effect of confinement by measuring the structural differences between water in contact with the top/bottom and outer/inner leaflets of both systems, using coherent second harmonic (SH) imaging (for the extended bilayer) and scattering (for the unilammelar liposomes). SH generation, a process in which two near infrared photons are combined into a visible photon with the double frequency (Figure 1A, energy level scheme). SH photons are generated by anisotropic distributions of molecules[46–53]. Isotropic bulk liquids do not generate coherent SH photons, and neither do fully symmetric systems. A single interface will change the isotropic distribution of water and induce SH contrast, while a bilayer with two identical leaflets facing opposite sides will not produce coherent SH photons. If the size of liposomes is too large to induce any type of spatial confinement in the inner shell, then the bilayer will have the same structural influence as an extended planar bilayer (and there will be interfacial coherent SH response).

Freestanding horizontal planar lipid membranes in water were formed following the Montal-Müller method[54–57]. Two separate lipid monolayer interfaces were formed at the air/water interface and apposed in an 80-120-µm-sized circular aperture in a 25-µm thick Teflon film, producing an extended planar bilayer with a radius of ~ 50 microns. This horizontally mounted membrane was imaged with a medium repetition rate, wide field nonlinear SH microscope[58]. Two counter-propagating 110 mW, 190 fs, 1032 nm, 200 kHz pulsed beams with an opening angle of 45° illuminate the membrane interface, such that phase matched SH photons are emitted along the surface normal (Figure 1A, see Ref.[59], and the materials and methods section). Figure 1B shows SH images recorded of (i) a pure water, i.e. without an interface, (ii) a symmetric lipid bilayer of identical lipid leaflets composed of 100 % DOPC (1,2-dioleoyl-sn-glycero-3-phosphocholine), and (iii) an asymmetric bilayer composed of 37.5:37.5:25 mol % 1,2-dipalmitoyl-sn-glycero-3-phosphoserine (DPPS), DOPC, and cholesterol (Chol) on the top leaflet, and 37.5:37.5:25 mol % 1,2-dipalmitoyl-sn-glycero-3-phosphocholine (DPPC), DOPC, and Chol on the bottom leaflet. Figure 1C shows a bar graph of the integrated counts of the corresponding SH images (i-iii) of bulk water and of the symmetric and asymmetric bilayers from 1B, normalized by the SSS response of bulk water[60]. SH Images (i) and (ii) produce a negligible number of photons, while SH image (iii) produces a significantly higher number of SH photons. The lack of a response from the symmetric bilayer agrees with earlier work[59,61] and confirms that two identical planar leaflets spaced 4 nm away and with opposite orientation cancel each other's SH response. In other words, they do not give rise to a difference in water structure. The coherent SH response of the asymmetric bilayer on the other hand does, because the structure on both leaflets is different. Here the SH signal arises from the charge-dipole interaction between the charged head groups and the dipolar water molecules, which creates a non-random orientational distribution of water dipoles along the surface normal.

Next, we investigate whether there are structural differences between water in contact with leaflets on either side of zwitterionic, net charge neutral DOPC liposomes, 58 – 140 nm in diameter. To probe the water structure in contact with these nanometric objects, we switch to second harmonic scattering (SHS)[62–69], as objects below the diffraction limit cannot be SH imaged[70]. Figure 1D shows an illustration of the SHS experiment. Figure 1E shows AR-SHS[60] patterns of 138 (69) nm and 72 (36) nm diameter (radius) liposomes made of neutral DOPC lipids. The patterns were recorded by focusing near-infrared 60 mW, 190 fs, 1032 nm, 200 kHz laser pulses in the liposome dispersion and detecting the emitted SH photons with a detector placed on a rotating arm. The emitted SH photons originate from both the liposome interfaces as well as from pure water. The coherent response from the aqueous interface is recorded with all beams polarized along the scattering plane (PPP polarization). It consists of the SH emission from the aqueous liposome dispersion from which the SH response of the neat water is subtracted.

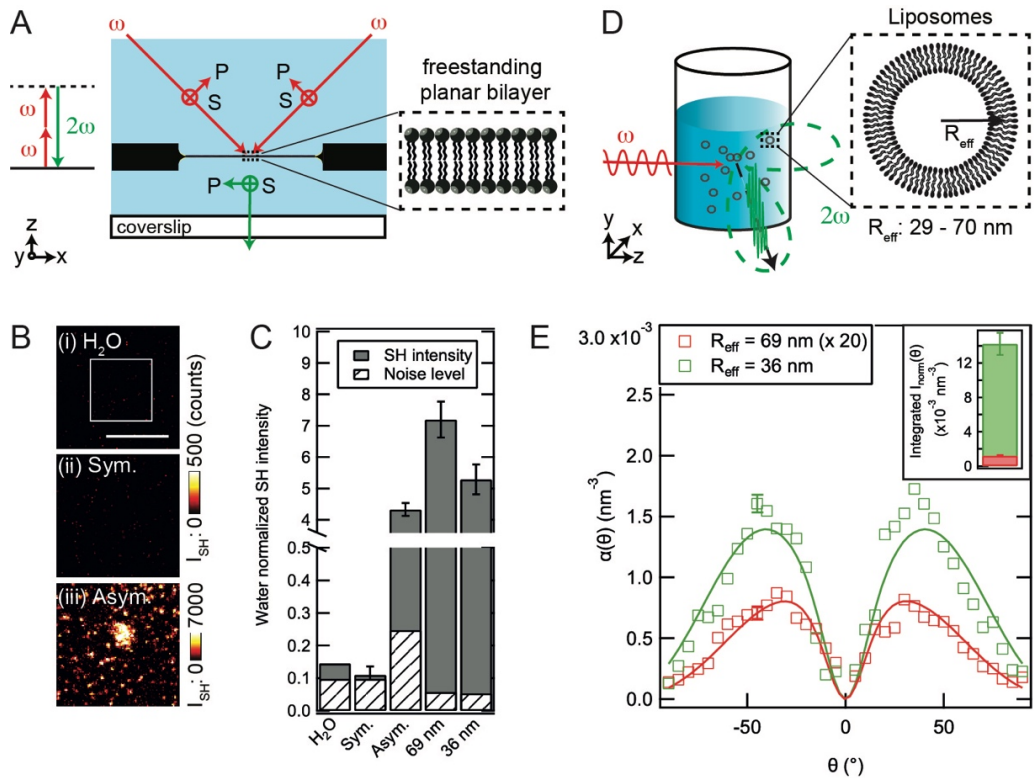

**Figure 1. Water structure around curved bilayers. (A)** Illustration of the SH imaging experiment. Two counter-propagating beams (190 fs, 1032 nm, ω, red arrows) overlap in space and time to illuminate a lipid bilayer membrane that is freely suspended in water. The bilayer is nearly planar and has a radius of curvature of ~ 50 μm, as dictated by the opening in which it is located. Non-resonant SH photons (2ω, green arrow) are collected (magnification: 50 x, NA = 0.65) in the phase-matched direction. All beams are polarized in the imaging plane. The energy level scheme is shown on the left. **(B)** SH images of (i) H$_2$O, (ii) a symmetric membrane composed of DOPC, (iii) an asymmetric membrane composed of 37.5:37.5:25 mol % DPPS:DOPC with cholesterol (top leaflet) and 37.5:37.5:25 mol % of DPPC:DOPC with cholesterol (bottom leaflet). The scale bar (20 μm) is the same for all images. The acquisition time is 560 ms for each image. **(C)** Bar graph of the power normalized SH intensity integrated over a 20 μm x 20 μm region of interest centered in the images (i), (ii), (iii). In the case of the neat water image, the normalized SH intensity was taken from 1 frame only. The SH intensities from panel (e) are also shown, prior to size normalization, on the right of the bar graph. The SH intensities of the images and scattering patterns were normalized to the SSS response of neat water in the same way. The dashed areas indicate the noise level for each measurement (computed as the standard deviation of the PPP response of neat water) also normalized by the SSS response of neat water. Note that the SS responses were recorded in the same away as the PPP responses. **(D)** Illustration of the AR-SHS scattering experiment performed on liposomes in solution. One beam (190-fs, 1032 nm, ω, red arrows) is focused into an aqueous liposome solution, stored in a cylindrical cuvette. SH photons (2ω, green arrow) are collected with a moveable detector in the (x,z) plane to record scattering patterns. All beams are polarized in the (x,z)-scattering plane (P polarization). **(E)** AR-SHS scattering patterns of DOPC liposomes in pure H$_2$O (0.5 mg/mL), recorded in PPP polarization combination. The error bar at -45° angle shows the maximum standard deviation over 20 measurements. The data are scaled to correct for differences in size distribution and number density of the scatterers (as described in the SI). The solid lines are fits extracted with the nonlinear light scattering algorithm, scaled accordingly. The inset shows the total value of $I_{norm}(\theta)$ integrated from -85 to +85° of scattering angle, for both patterns.

Just as with the SH images, this differential intensity was divided by the incoherent isotropic SH scattering of water recorded in the SSS polarization combination under the same illumination conditions[60] (i.e. all beams polarized perpendicular to the scattering plane). The resulting SH response, $S(\theta)$ (eq. S1 in the S1) thus reports of the intensity of the liposomes relative to that of isotropic bulk water in the same volume. Since the scattered intensity depends both on the size of the liposome and the number of liposomes in solution, we computed a size and number density independent term $\alpha(\theta)$, which is plotted in Figure 1E. To obtain $\alpha(\theta)$, $S(\theta)$ was normalized to both the number of liposomes in solution (Eq. S1) as well as the size of the liposomes, using an approximate intensity dependence[71] on the radius $R$ of $R^6$ (Eq. S8). Because the liposomes in the solution do not all have a single diameter but a size distribution we calculated an effective radius ($R_{eff}$) that generates the same dynamic light scattering pattern as the distribution of a sample that has multiple radii in the distribution[72] (see SI S6). $R_{eff}$ is used to determine the size normalized single liposome SH intensity ($\alpha(\theta)$). The fits in Figure 1E were obtained using nonlinear light scattering theory as described below. The inset shows the total value of $\alpha(\theta)$ integrated from -85 to +85° of scattering angle, for both patterns.

Figure 1E shows that SH intensity is visible for both liposome samples, indicating that the inner leaflets water structure differs from the outer leaflets water structure. The inset shows the angle integrated SH intensities. The smaller sized liposomes generate a ~ 40 times larger SH response per object. Considering that the SH photons derive from the difference in orientational distribution of water between the inner and outer leaflet, going from 69 nm to 36 nm in effective radius, results in a relative change in unmatched interfacial area of a factor of 2, which should result in a difference in $\alpha(\theta)$ of 4. The found value of ~40 therefore indicates that the difference in water structure is much bigger for smaller liposomes than would have logically been expected. It suggests that confinement effects are already apparent for vesicles in $H_2O$ with diameters as large as 140 nm, which is remarkably larger than previously reported[10,14,20,24]. This result warrants a more in-depth size analysis.

**3D confinement restricts the hydrogen bond network of $H_2O$ but not $D_2O$.** To investigate this observation further, we repeated the measurements of Figure 1E using DOPC liposomes having effective radii between 29 nm to 70 nm. Liposomes were prepared in both $H_2O$ and $D_2O$ and AR-SHS patterns were recorded in two different polarization combinations (PPP, with all beams polarized in the horizontal scattering plane, and PSS, with the SH beam polarized along the scattering plane, and the incoming beams polarized vertical to the scattering plane. The measured $S(\theta)$ patterns were fitted using nonlinear light scattering theory to fully account for the size dependence. The application of nonlinear light scattering theory to SHS has been treated in detail in Ref.[73] and was applied to determine the surface potential of charged liposomes[60,74], silica[75] and titania[76] particles in aqueous solution. The Rayleigh-Gans-Debye nonlinear light scattering theory generates solutions to the Maxwell equations that describe the light-matter interaction of the SHS process. Based on these solutions, for single scattering objects with a radius < ~ 250 nm the non-resonant SH scattered intensity is determined by just two parameters: The difference between the anisotropic orientational distributions of water molecules adjacent to the inner and outer leaflets, represented by $\chi_{s,2}^{(2)}$, and the electrostatic surface potential difference ($\Phi_0$) between the inner and outer leaflet. To describe the recorded AR-SHS patterns (Figure S1), we used the expressions given in Ref.[74] (section S4 of the SI) and used the parameters as provided in S4 and Tables S1 and S2. Since DOPC membranes are zwitterionic, and therefore charge neutral, the differential surface potential ($\Phi_0$) was set to $\Phi_0$=0 mV, and the fits to the AR-SHS patterns were made using only $\chi_{s,2}^{(2)}$ as a single fit parameter, $\chi_{s,2}^{(2)}$. It represents the net amount of oriented water, as induced by any non-electrostatic interaction.

Figure 2A shows the extracted $\chi_{s,2}^{(2)}$ values from the polarimetric AR-SHS measurements of neutral DOPC liposomes dispersed in light (red filled circles) and heavy (black filled squares) water as a function of the effective radius. For comparison, Figure 2B shows the values of the corresponding integrated scattering patterns recorded in the PPP polarization of pure bulk water (without liposomes). The intensity of bulk $D_2O$ is bigger than the one of bulk $H_2O$ by a factor of 1.95, which means that $D_2O$ molecules have an on average larger molecular hyperpolarizability than $H_2O$ – in agreement with previous work[77].

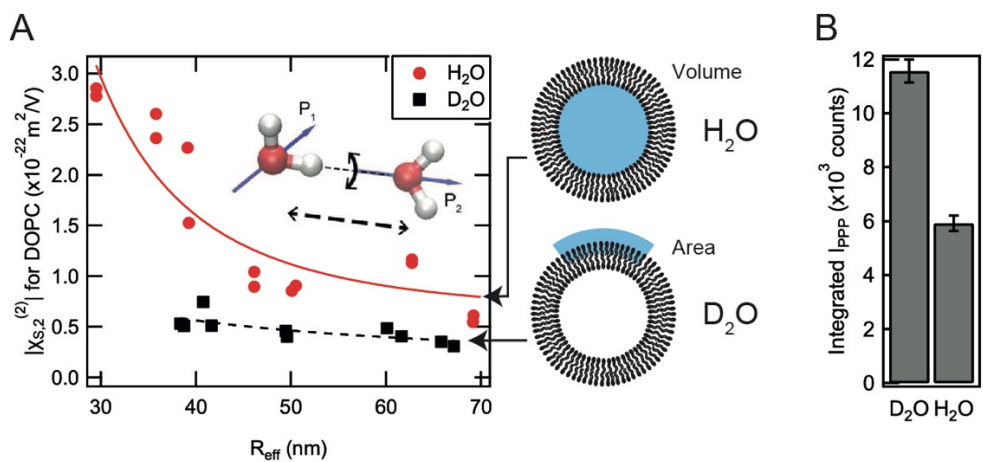

**Figure 2. 3D confinement in light and heavy water. (A)** Absolute value of $\chi_{s,2}^{(2)}$ computed for DOPC liposomes in $H_2O$ (red filled circles) and $D_2O$ (black filled squares), as a function of effective radius. The lines represent fits for $R_{eff}^{-3}$ (continuous, red) and $R_{eff}^{-1}$ (dashed, black) dependences, respectively. The inset inside the graph shows two hydrogen-

bonded water molecules, highlighting the hydrogen bond stretch and bend modes that result in breaking of the H-bond. The illustrations on the right-hand side of the graph illustrate the origin of the generated intensity. **(B)** The integrated SH intensity of neat water in PPP polarization, integrated over the angular range $-85° < \theta < 85°$ for $H_2O$ and $D_2O$, showing that the bulk response of $D_2O$ is bigger than that of $H_2O$.

Turning now to the size dependencies of Figure 2A. Both light and heavy water show a size dependence of the single liposome intensity as a function of size. The liposome induced change in orientational order is ~ 8 x smaller for $D_2O$ compared to $H_2O$. As noted above, there is an obvious reason why SH intensity changes are generated and why the single liposome response is size dependent. Since the inner leaflet has a smaller surface area than the outer leaflet (because the membrane is ~4 nm thick), there is always a difference in the number of interfacial molecules. Because this difference in thickness does not depend on the radius, reducing the size will increase the relative difference in interfacial area between the inner and outer leaflet. This type of differential SH response scales with the inverse of the radius so that $\chi_{s,2}^{(2)} \sim \frac{1}{R_{eff}}$. The black line in Figure 2A represents this $1/R_{eff}$ dependence and is in good agreement with the $D_2O$ data. For $H_2O$, it does not match with the increase in the $\chi_{s,2}^{(2)}$ magnitude with reducing $R_{eff}$ values, which is much steeper. This means that the orientational distribution of light water changes much more than can be explained by an increase in relative interfacial area when going to smaller sizes.

3D confinement of water forms an explanation: As the shell of charge-neutral lipids hinders subtle longer-ranged structural arrangements it changes the inner water structure compared to the outer water structure. If the confinement by the 3D membrane results in a H-bond network that has a slightly different structure from bulk water, the difference in SH response is expected to scale with the inner volume of the liposome and one expects that $\chi_{s,2}^{(2)}(R_{eff}) \sim \frac{1}{R_{eff}^3}$. Figure 2A shows that the $H_2O$ $\chi_{s,2}^{(2)}$ data can indeed be captured by a $\frac{1}{R_{eff}^3}$ dependence, represented by the red line, which means that $H_2O$ has subtle interactions in the H-bond network that extend over distances in the size range up to 140 nm. While the $D_2O$ size dependence data can be explained by a change in relative interfacial area, as one might expect from simple geometrical arguments, for $H_2O$, we observe a very different dependence that scales with the available inner liposome volume.

Thus, the orientational distribution function of $H_2O$ is more dramatically changed by a restriction in the degrees of freedom of the H-bond network induced by 3D spatial confinement. In this study the largest liposomes with $R_{eff}$ ~70 nm contain ~4×10$^7$ molecules, and are ~518 water molecules in diameter. The smallest liposomes with $R_{eff}$ ~29 nm contain ~3×10$^6$ molecules and span ~215 water molecules in diameter. Our results show that spatially confining light water to a volume holding >10$^7$ water molecules can already considerably impact the configuration of the 3D H-bond network of $H_2O$. For $D_2O$, however, the H-bond network is not changing its configuration beyond our detection limit, and the 3D confinement limit should be <10$^6$ water molecules. This means that the spatial extent over which $H_2O$ molecules influence each other is much larger than the extent over which $D_2O$ molecules influence each other.

Comparing light and heavy water, $D_2O$ has a slightly different hydrogen bond network compared to $H_2O$. Although the net hydrogen bond strength differs by only a few percent at most[78] in strength, the different degrees of freedom that together determine the H-bond network reorganization have different strengths. Breaking and reforming hydrogen bonds occurs through both a stretching and a rotational (or librational) mode[79–81], as illustrated in the inset of Figure 2A. The H-bond stretching mode is stronger for $H_2O$ and weaker for $D_2O$, while the H-bond rotational motion is weaker for $H_2O$ and stronger for $D_2O$. Overall $D_2O$ has slightly stronger H-bonds, leading to an increase in melting point of 3.8 °C at ambient pressure, (and 7°C for the temperature of highest density). Therefore, if $H_2O$ and $D_2O$ display a different response to the 3D confinement experiment of Figure 2, the origin of the size dependence arises from 3D H-bond restructuring invoked by a limitation in the spatial degrees of freedom. Identical liposome induced changes in the orientational distributions of water molecules in light and heavy water would have resulted in an ~ 8 x bigger induced change in $\chi^{(2)}_{s,2}$ for $D_2O$, as well as an identical size dependence for $D_2O$ and $H_2O$. Since the data in Figure 2 shows exactly the opposite with $\chi^{(2)}_{s,2}$ being bigger for the same size of liposome in $H_2O$ than in $D_2O$, a different radial dependence (Figure 2a), and a larger overall SH intensity of bulk $D_2O$ (Figure 2b), we conclude that the H-bond networks are not only subtly different on sub-nanometric length scales but also on length scales up to ~ 100 nm. Our finding is qualitative in agreement with previous work: $H_2O$ solvates more easily than $D_2O$[82], the H-bond network of $H_2O$ is more easily perturbed by electrolytes[77], or temperature changes[83] compared to that of $D_2O$. Surface tension anomalies induced by salt ions, known as the Jones-Ray effect, also occur at lower concentrations $H_2O$ than in $D_2O$[84]. Based on the latter three studies that showed a difference on the order of 6-7 in the electrolyte concentrations necessary to perturb the bulk structure of water, we expect confinement effects to show up in $D_2O$ at a length scale of ~140 / 7 = 20 nm (diameter), which is below the signal to noise level of our experiment.

The length scale over which the confinement effects are observed in $H_2O$ in this study are much larger than the confinement length scales that have been previously observed, for example in porous silica[11,24] (~10 nm), at surfaces with force measurements[26–28] (~4 nm), within reverse micelles[14,17] (~5 nm), thin channels[25] (~20 nm), or carbon nanotubes[10,19] (~2 nm). Based on the above explanation, that involves a restriction in the distribution of heterogeneous substructures, the difference is explained in the available degrees of freedom: a nanopore or a slit is in contact with surrounding bulk liquid, just like the water in a porous material and the water between an AFM tip and a substrate or the two substrates of a surface force apparatus[85]. Given that the reorientation time scale of water and the lifetime of wires and cavities are on the order of picoseconds, there is enough contact with a reservoir of bulk water to enable a relaxation of the restriction. For the 3D liposome shells as investigated here this is not the case, since the hydrogen bond networks of the water molecules inside the shell are not connected to the hydrogen bond network outside the shell.

**Conclusions**

Summarizing, the 3D confinement of water was investigated by performing second harmonic imaging and scattering experiments on lipid bilayer membranes having near-infinite curvature, and radii between 29 and 70 nm. These charged-neutral zwitterionic liposomes contain ~$3\times10^6$ - $4\times10^7$ water molecules. We measured the change in water structure between the inside and outside water of the liposomes and compared it to the structural differences found for extended planar lipid membranes. While the latter show no detectable coherent SH response, the liposomes show a size dependent change in the SH intensity that is different for light and heavy water. The membrane induced change in the order of water was retrieved from nonlinear light scattering modeling, and normalized to the size and number density, so that size dependent single liposome responses were obtained. For heavy water the shell-size normalized measured change in water ordering matches the change in available interfacial area. For light water, however, the change in orientational order is 8 x larger compared to heavy water and scales with the inner volume of the shells. Our findings demonstrate that 3D structural confinement effects in water can therefore involve on the order of $10^7$ water molecules, and that there is a remarkably large difference between $H_2O$ and $D_2O$. That this has not been found so far is due to two things: the unprecedented sensitivity of femtosecond second harmonic scattering to instantaneous longer ranged structural changes[77] and the fact that previous measurements have studied systems in which the confined water was structurally in contact with an open bath. The presented data demonstrates the intricate nature of water, and the need to investigate water-water interactions over longer distances which clearly does play a role in nano- and microscopic structuring of aqueous systems. Biochemical equilibria that determine the aqueous cellular environment typically depend on small changes in hydration structure and energy landscapes. They might well be impacted by the observed confinement effects.

**Experimental methods**

The materials and methods, and the details of the second harmonic imaging and the angle-resolved second harmonics scattering experiments are described in the Supporting Information.

**Acknowledgments**

This work was supported by the Julia Jacobi Foundation, and the European Research Council grant 616305.

**Supporting Information Available**

Materials and Methods (S1), Second harmonic imaging (S2), Angle-resolved second harmonic scattering (S3), AR-SHS Theory (S4), Fits to second harmonic scattering theory (S5), Determination of effective radius (S6).

**Author contributions**

N.D., O.B.T. and C. L. performed experiments, N.D., O.B.T, and S.R. interpreted the experimental data. S.R., O.B.T., and N.D. wrote the manuscript. S.R. conceived and supervised the work.

**Competing interests**

The authors declare no competing interests.

# References

1. Ball, P. Water as an active constituent in cell biology. *Chem. Rev.* **108**, 74–108 (2008).

2. Johnson, C. M. & Baldelli, S. Vibrational Sum Frequency Spectroscopy Studies of the Influence of Solutes and Phospholipids at Vapor/Water Interfaces Relevant to Biological and Environmental Systems. *Chem. Rev.* **114**, 8416–8446 (2014).

3. Tarun, O. B., Eremchev, M. Yu., Radenovic, A. & Roke, S. Spatiotemporal Imaging of Water in Operating Voltage-Gated Ion Channels Reveals the Slow Motion of Interfacial Ions. *Nano Lett.* **19**, 7608–7613 (2019).

4. Vanselous, H. & Petersen, P. B. Between a rock and a soft place. *Nature Chemistry* **8**, 527–528 (2016).

5. Laage, D., Elsaesser, T. & Hynes, J. T. Water Dynamics in the Hydration Shells of Biomolecules. *Chem. Rev.* **117**, 10694–10725 (2017).

6. Bellissent-Funel, M.-C. *et al.* Water Determines the Structure and Dynamics of Proteins. *Chem. Rev.* **116**, 7673–7697 (2016).

7. Yan, E. C. Y., Fu, L., Wang, Z. & Liu, W. Biological Macromolecules at Interfaces Probed by Chiral Vibrational Sum Frequency Generation Spectroscopy. *Chem. Rev.* **114**, 8471–8498 (2014).

8. McDermott, M. L., Vanselous, H., Corcelli, S. A. & Petersen, P. B. DNA's Chiral Spine of Hydration. *ACS Cent. Sci.* **3**, 708–714 (2017).

9. Kocsis, I. *et al.* Oriented chiral water wires in artificial transmembrane channels. *Sci Adv* **4**, eaao5603 (2018).

10. Köhler, M. H., Bordin, J. R., de Matos, C. F. & Barbosa, M. C. Water in nanotubes: The surface effect. *Chemical Engineering Science* **203**, 54–67 (2019).

11. Knight, A. W., Kalugin, N. G., Coker, E. & Ilgen, A. G. Water properties under nano-scale confinement. *Scientific Reports* **9**, 8246 (2019).

12. Gao, Z., Giovambattista, N. & Sahin, O. Phase Diagram of Water Confined by Graphene. *Scientific Reports* **8**, 6228 (2018).

13. Cicero, G., Grossman, J. C., Schwegler, E., Gygi, F. & Galli, G. Water Confined in Nanotubes and between Graphene Sheets: A First Principle Study. *J. Am. Chem. Soc.* **130**, 1871–1878 (2008).

14. Fayer, M. & Levinger, N. E. Analysis of Water in Confined Geometries and at Interfaces. *Annual review of analytical chemistry* **3**, 89–107 (2010).

15. Levinger, N. E. Water in Confinement. *Science* **298**, 1722 (2002).

16. Fayer, M. D. Water in a Crowd. *Physiology* **26**, 381–392 (2011).

17. Dokter Adriaan M., Woutersen Sander, & Bakker Huib J. Inhomogeneous dynamics in confined water nanodroplets. *Proceedings of the National Academy of Sciences* **103**, 15355–

15358 (2006).

18.     Baksi, A., Ghorai, P. Kr. & Biswas, R. Dynamic Susceptibility and Structural Heterogeneity of Large Reverse Micellar Water: An Examination of the Core–Shell Model via Probing the Layer-wise Features. *J. Phys. Chem. B* **124**, 2848–2863 (2020).

19.     Hassan, J., Diamantopoulos, G., Homouz, D. & Papavassiliou, G. Water inside carbon nanotubes: structure and dynamics. *Nanotechnology Reviews* **5**, 341–354 (2016).

20.     Tsimpanogiannis, I. N. *et al.* Self-diffusion coefficient of bulk and confined water: a critical review of classical molecular simulation studies. *Molecular Simulation* **45**, 425–453 (2019).

21.     Chakraborty, S., Kumar, H., Dasgupta, C. & Maiti, P. K. Confined Water: Structure, Dynamics, and Thermodynamics. *Acc. Chem. Res.* **50**, 2139–2146 (2017).

22.     Klameth, F. & Vogel, M. Structure and dynamics of supercooled water in neutral confinements. *J. Chem. Phys.* **138**, 134503 (2013).

23.     Harrach, M. F., Klameth, F., Drossel, B. & Vogel, M. Effect of the hydroaffinity and topology of pore walls on the structure and dynamics of confined water. *J. Chem. Phys.* **142**, 034703 (2015).

24.     Burris, P. C., Laage, D. & Thompson, W. H. Simulations of the infrared, Raman, and 2D-IR photon echo spectra of water in nanoscale silica pores. *J. Chem. Phys.* **144**, 194709 (2016).

25.     Fumagalli, L. *et al.* Anomalously low dielectric constant of confined water. *Science* **360**, 1339 (2018).

26.     Li, T.-D., Gao, J., Szoszkiewicz, R., Landman, U. & Riedo, E. Structured and viscous water in subnanometer gaps. *Phys. Rev. B* **75**, 115415 (2007).

27.     Li, T. D. & Riedo, E. Nonlinear Viscoelastic Dynamics of Nanoconfined Wetting Liquids. *Phys. Rev. Lett.* **100**, 106102–106102 (2008).

28.     Watkins, M., Berkowitz, M. L. & Shluger, A. L. Role of water in atomic resolution AFM in solutions. *Phys. Chem. Chem. Phys.* **13**, 12584–12594 (2011).

29.     Chaplin, M. F. Structuring and Behaviour of Water in Nanochannels and Confined Spaces. in *Adsorption and Phase Behaviour in Nanochannels and Nanotubes* (eds. Dunne, L. J. & Manos, G.) 241–255 (Springer Netherlands, 2010). doi:10.1007/978-90-481-2481-7_11.

30.     Ayotte, P. *et al.* Infrared spectroscopy of negatively charged water clusters: Evidence for a linear network. *J. Chem. Phys.* **110**, 6268–6277 (1999).

31.     Robertson, W. H., Diken, E. G., Price, E. A., Shin, J. W. & Johnson, M. A. Spectroscopic determination of the OH- solvation shell in the OH-...(H2O)(n) clusters. *Science* **299**, 1367–1372 (2003).

32.     Fioretto, D., Freda, M., Mannaioli, S., Onori, G. & Santucci, A. Infrared and Dielectric Study of Ca(AOT)2 Reverse Micelles. *J. Phys. Chem. B* **103**, 2631–2635 (1999).

33.     Patra, A., Luong, T. Q., Mitra, R. K. & Havenith, M. The influence of charge on the structure and dynamics of water encapsulated in reverse micelles. *Phys. Chem. Chem. Phys.*

**16**, 12875–12883 (2014).

34. Cole, W. T. S. *et al.* Terahertz VRT Spectroscopy of the Water Hexamer-h12 Cage: Dramatic Libration-Induced Enhancement of Hydrogen Bond Tunneling Dynamics. *J. Phys. Chem. A* **122**, 7421–7426 (2018).

35. Gallo, P. *et al.* Water: A Tale of Two Liquids. *Chem. Rev.* **116**, 7463–7500 (2016).

36. Zhong, D., Pal, S. K. & Zewail, A. H. Biological water: A critique. *Chemical Physics Letters* **503**, 1–11 (2011).

37. Richard, T., Mercury, L., Poulet, F. & d'Hendecourt, L. Diffuse reflectance infrared Fourier transform spectroscopy as a tool to characterise water in adsorption/confinement situations. *Journal of Colloid and Interface Science* **304**, 125–136 (2006).

38. Byl, O. *et al.* Unusual Hydrogen Bonding in Water-Filled Carbon Nanotubes. *J. Am. Chem. Soc.* **128**, 12090–12097 (2006).

39. Cerveny, S., Mallamace, F., Swenson, J., Vogel, M. & Xu, L. Confined Water as Model of Supercooled Water. *Chem. Rev.* **116**, 7608–7625 (2016).

40. Tsukahara, T., Hibara, A., Ikeda, Y. & Kitamori, T. NMR Study of Water Molecules Confined in Extended Nanospaces. *Angewandte Chemie International Edition* **46**, 1180–1183 (2007).

41. Fritsch, S., Potestio, R., Donadio, D. & Kremer, K. Nuclear Quantum Effects in Water: A Multiscale Study. *J. Chem. Theory Comput.* **10**, 816–824 (2014).

42. Coudert, F.-X., Vuilleumier, R. & Boutin, A. Dipole Moment, Hydrogen Bonding and IR Spectrum of Confined Water. *ChemPhysChem* **7**, 2464–2467 (2006).

43. Muñoz-Santiburcio, D., Wittekindt, C. & Marx, D. Nanoconfinement effects on hydrated excess protons in layered materials. *Nature Communications* **4**, 2349 (2013).

44. Schlaich, A., Knapp, E. W. & Netz, R. R. Water Dielectric Effects in Planar Confinement. *Phys. Rev. Lett.* **117**, 048001 (2016).

45. Schürer, B., Wunderlich, S., Sauerbeck, C., Peschel, U. & Peukert, W. Probing colloidal interfaces by angle-resolved second harmonic light scattering. *Phys. Rev. B* **82**, 241404–241404 (2010).

46. Bloembergen, N. Surface nonlinear optics: a historical overview. *Appl. Phys. B* **68**, 289–293 (1999).

47. Eisenthal, K. B. Second harmonic spectroscopy of aqueous nano- and microparticle interfaces. *Chem. Rev.* **106**, 1462–1477 (2006).

48. Shen, Y. R. Optical Second Harmonic Generation at Interfaces. *Ann. Rev. Phys. Chem.* **40**, 327–350 (1989).

49. Shen, Y. R. *The principles of nonlinear optics*. (Wiley, 1984).

50. Shen, Y. R. Surface properties probed by second-harmonic and sum-frequency generation. *Nature* **337**, 519–519 (1989).

51. Geiger, F. M. Second Harmonic Generation, Sum Frequency Generation, and chi((3)): Dissecting Environmental Interfaces with a Nonlinear Optical Swiss Army Knife. *Annu. Rev. Phys. Chem.* **60**, 61–83 (2009).

52. Azam, Md. S. & Gibbs-Davis, J. M. Monitoring DNA Hybridization and Thermal Dissociation at the Silica/Water Interface Using Resonantly Enhanced Second Harmonic Generation Spectroscopy. *Anal. Chem.* **85**, 8031–8038 (2013).

53. Rehl, B. *et al.* New Insights into Chi(3) Measurements: Comparing Nonresonant Second Harmonic Generation and Resonant Sum Frequency Generation at the Silica/Aqueous Electrolyte Interface. *J. Phys. Chem. C* **123**, 10991–11000 (2019).

54. Montal, M. & Mueller, P. Formation of Bimolecular Membranes from Lipid Monolayers and a Study of Their Electrical Properties. *Proc Natl Acad Sci USA* **69**, 3561 (1972).

55. Krylov, A. V., Pohl, P., Zeidel, M. L. & Hill, W. G. Water Permeability of Asymmetric Planar Lipid Bilayers: Leaflets of Different Composition Offer Independent and Additive Resistances to Permeation. *Journal of General Physiology* **118**, 333–340 (2001).

56. Horner, A., Akimov, S. A. & Pohl, P. Long and Short Lipid Molecules Experience the Same Interleaflet Drag in Lipid Bilayers. *Phys. Rev. Lett.* **110**, 268101 (2013).

57. White, S. H., Petersen, D. C., Simon, S. & Yafuso, M. Formation of planar bilayer membranes from lipid monolayers. A critique. *Biophysical journal* **16**, 481–489 (1976).

58. Macias-Romero, C. *et al.* High throughput second harmonic imaging for label-free biological applications. *Opt. Express* **22**, 31102–31112 (2014).

59. Tarun, O. B., Hannesschläger, C., Pohl, P. & Roke, S. Label-free and charge-sensitive dynamic imaging of lipid membrane hydration on millisecond time scales. *Proc Natl Acad Sci USA* **115**, 4081 (2018).

60. Lütgebaucks, C., Gonella, G. & Roke, S. Optical label-free and model-free probe of the surface potential of nanoscale and microscopic objects in aqueous solution. *Phys. Rev. B* **94**, 195410 (2016).

61. Ries, R. S., Choi, H., Blunck, R., Bezanilla, F. & Heath, J. R. Black Lipid Membranes: Visualizing the Structure, Dynamics, and Substrate Dependence of Membranes. *J. Phys. Chem. B* **108**, 16040–16049 (2004).

62. Eisenthal, K. B. Liquid Interfaces Probed by Second-Harmonic and Sum-Frequency Spectroscopy. *Chem Rev* **96**, 1343–1360 (1996).

63. Liu, Y., Yan, E. C. Y. & Eisenthal, K. B. Effects of bilayer surface charge density on molecular adsorption and transport across liposome bilayers. *Biophysical Journal* **80**, 1004–1012 (2001).

64. Liu, J., Subir, M., Nguyen, K. & Eisenthal, K. B. Second Harmonic Studies of Ions Crossing Liposome Membranes in Real Time. *J. Phys. Chem. B* **112**, 15263–15266 (2008).

65. Gonella, G., Gan, W., Xu, B. & Dai, H.-L. The Effect of Composition, Morphology, and Susceptibility on Nonlinear Light Scattering from Metallic and Dielectric Nanoparticles. *J. Phys. Chem. Lett.* **3**, 2877–2881 (2012).

66. Schneider, L., Schmid, H. J. & Peukert, W. Influence of particle size and concentration on the second-harmonic signal generated at colloidal surfaces. *Appl. Phys. B* **87**, 333–339 (2007).

67. Sharifian Gh., M., Wilhelm, M. J. & Dai, H.-L. Label-Free Optical Method for Quantifying Molecular Transport Across Cellular Membranes In Vitro. *J. Phys. Chem. Lett.* **7**, 3406–3411 (2016).

68. Wilhelm, M. J., Sharifian Gh., M. & Dai, H.-L. Influence of molecular structure on passive membrane transport: A case study by second harmonic light scattering. *J. Chem. Phys.* **150**, 104705 (2019).

69. Zeng, J., Eckenrode, H. M., Dai, H.-L. & Wilhelm, M. J. Adsorption and transport of charged vs. neutral hydrophobic molecules at the membrane of murine erythroleukemia (MEL) cells. *Colloids and Surfaces B: Biointerfaces* **127**, 122–129 (2015).

70. Barry, R. & Masters, R. *Handbook of Biomedical Nonlinear Optical Microscopy*. (Published by Oxford University Press, 2008).

71. de Beer, A. G. F. & Roke, S. Nonlinear Mie theory for second-harmonic and sum-frequency scattering. *Phys. Rev. B* **79**, 155420–155429 (2009).

72. Smolentsev, N., Lütgebaucks, C., Okur, H. I., de Beer, A. G. F. & Roke, S. Intermolecular Headgroup Interaction and Hydration as Driving Forces for Lipid Transmembrane Asymmetry. *J. Am. Chem. Soc.* **138**, 4053–4060 (2016).

73. Gonella, G., Lütgebaucks, C., de Beer, A. G. F. & Roke, S. Second Harmonic and Sum-Frequency Generation from Aqueous Interfaces Is Modulated by Interference. *J. Phys. Chem. C* **120**, 9165–9173 (2016).

74. Lütgebaucks, C., Macias-Romero, C. & Roke, S. Characterization of the interface of binary mixed DOPC:DOPS liposomes in water: The impact of charge condensation. *J. Chem. Phys.* **146**, 044701 (2017).

75. Marchioro, A. *et al.* Surface Characterization of Colloidal Silica Nanoparticles by Second Harmonic Scattering: Quantifying the Surface Potential and Interfacial Water Order. *The Journal of Physical Chemistry C* **123**, 20393–20404 (2019).

76. Bischoff, M., Biriukov, D., Předota, M., Roke, S. & Marchioro, A. Surface Potential and Interfacial Water Order at the Amorphous $TiO_2$ Nanoparticle/Aqueous Interface. *J. Phys. Chem. C* **124**, 10961–10974 (2020).

77. Chen, Y. *et al.* Electrolytes induce long-range orientational order and free energy changes in the H-bond network of bulk water. *Sci. Adv.* **2**, e1501891 (2016).

78. Soper, A. & Benmore, C. Quantum Differences between Heavy and Light Water. *Phys. Rev. Lett.* **101**, 065502 (2008).

79. Habershon, S., Markland, T. E. & Manolopoulos, D. E. Competing quantum effects in the dynamics of a flexible water model. *J Chem Phys* **131**, 024501 (2009).

80. Li, X. Z., Walker, B. & Michaelides, A. Quantum nature of the hydrogen bond. *Proc. Natl. Acad. Sci. U.S.A.* **108**, 6369–6373 (2011).

81. Berendsen, H. J. C., Postma, J. P. M., van Gunsteren, W. F., DiNola, A. & Haak, J. R. Molecular dynamics with coupling to an external bath. *J. Chem. Phys.* **81**, 3684–3690 (1984).

82. Marcus, Y. & Ben-Naim, A. A study of the structure of water and its dependence on solutes, based on the isotope effects on solvation thermodynamics in water. *J. Chem. Phys.* **83**, 4744–4759 (1985).

83. Chen, Y., Dupertuis, N., Okur, H. I. & Roke, S. Temperature dependence of water-water and ion-water correlations in bulk water and electrolyte solutions probed by femtosecond elastic second harmonic scattering. *J. Chem. Phys.* **148**, 222835 (2018).

84. Okur, H. I., Chen, Y., Wilkins, D. M. & Roke, S. The Jones-Ray effect reinterpreted: Surface tension minima of low ionic strength electrolyte solutions are caused by electric field induced water-water correlations. *Chem. Phys. Lett.* **684**, 433–442 (2017).

85. Tivony, R., Yaakov, D. B., Silbert, G. & Klein, J. Direct Observation of Confinement-Induced Charge Inversion at a Metal Surface. *Langmuir* **31**, 12845–12849 (2015).